\documentclass[lettersize,journal]{IEEEtran}
\usepackage{cite}
\usepackage{amsmath}
\usepackage{amssymb}
\usepackage{algorithmic}
\usepackage{array}
\usepackage{graphicx}
\usepackage{bm}
\usepackage[caption=false,font=normalsize,labelfont=sf,textfont=sf]{subfig}
\usepackage{subfig}
\usepackage{url}
\usepackage{xcolor}

\begin{document}
\title{DRL-Based Joint Beamforming and Surface Shape Optimization for Flexible Intelligent Metasurface-Aided ISAC Systems
}

\author{Maoyuan Wang,
        Qian Zhang,~\IEEEmembership{Graduate Student Member,~IEEE},
        Jiancheng An,~\IEEEmembership{Senior Member,~IEEE},
        Xuejun Cheng,
        Zheng Dong,~\IEEEmembership{Member,~IEEE},
        and Deqiang Wang,~\IEEEmembership{Senior Member,~IEEE}

		\thanks{This research was supported in part by the Shandong Provincial Natural Science Foundation under Grant ZR2023LZH003 and the National Key R\&D Program of China under Grant 2024YFF0727101. The work of J. An was supported by the National Natural Science Foundation of China (NSFC) under Grant 62471096. The scientific calculations in this paper have been done on the HPC Cloud Platform of Shandong University.}
		\thanks{Maoyuan Wang, Xuejun Cheng, Zheng Dong, and Deqiang Wang are with the School of Information Science and Engineering, Shandong University, Qingdao 266237, China (e-mail: \{maoyuanwang2024, chengxuejun\}@mail.sdu.edu.cn; \{zhengdong, wdq\_sdu\}@sdu.edu.cn).}
		\thanks{Qian Zhang is with the School of Computer and Communication Engineering, Northeastern University at Qinhuangdao, Qinhuangdao 066004, China, and with the School of Information Science and Engineering, Shandong University, Qingdao 266237, China (e-mail: qianzhang2021@mail.sdu.edu.cn).}
		\thanks{J. An is with the School of Electronic Science and Engineering, University of Electronic Science and Technology of China (UESTC), Chengdu, 611731, China (e-mail: jiancheng.an@uestc.edu.cn).}
		\thanks{Corresponding author: Zheng Dong.}
     
}

\maketitle
\begin{abstract}
Integrated sensing and communication (ISAC) unifies high-precision sensing and wireless data transmission. In this paper, we investigate the design of ISAC systems enabled by flexible intelligent metasurface (FIM) and aim to minimize the Cramér–Rao bound (CRB) with quality of service (QoS) constraints using deep reinforcement learning (DRL). Specifically, we formulate the joint design of beamforming matrix and FIMs surface shape to reduce the CRB subject to transmit power, QoS and the FIMs surface shape constraints. However, the non-convex formulation makes optimization problem difficult to solve. To tackle this issue, we develop a deep deterministic policy gradient (DDPG) actor critic DRL scheme for the joint design, guided by a constraint aware reward to progressively improve sensing performance. Numerical results demonstrate that jointly optimizing the beamforming matrix and the FIMs surface shape substantially decreases CRB while ensuring communication quality compared with existing rigid arrays.
\end{abstract}


\section{Introduction}

\IEEEPARstart{I}{n} the future sixth generation (6G) era, wireless communications technologies are expected to support a broad range of emerging vertical applications such as urban digital twins, smart factories, and autonomous vehicles that require both ultra-reliable, low-latency communication and high-fidelity sensing for perception, localization, and tracking~\cite{dong2025communication}. To satisfy these requirements while alleviating spectrum scarcity and reducing system cost, communication systems evolve from supporting just communication to joint sensing and communications, where multi-antenna beamforming is a key enabler, as it focuses transmit energy and separating spatial channels to support joint sensing and communication. Integrated sensing and communications (ISAC) systems that exploit multi-antenna processing emerge as an important direction that unifies wireless communications and radar sensing for efficient spectrum and hardware utilization~\cite{zhang2026cram,xiu2026robust2,liu2022integrated,gonzalez2024integrated,xiu2025movable2}.

Additionally, metasurfaces play a crucial role in 6G networks, and a primary application is the deployment of reconfigurable intelligent surfaces (RISs)~\cite{zhang2025multi}. By adaptively changing the phase of incident electromagnetic waves, RIS can extend coverage, and increase data rates~\cite{zhang2023robust,zhang2024practical,pan2022overview}. Furthermore, in~\cite{yang2025secure}, Yang \textit{et al.} investigated a RIS-assisted ISAC system and jointly optimized the transmit beamforming of the base station (BS) and the RIS phase shifts with the objective of minimizing the Cramér–Rao bound (CRB). In~\cite{peng2024cramer}, Peng \textit{et al.} proposed a simultaneously transmitting and reflecting reconfigurable intelligent surface (STAR-RIS) enabled ISAC system that minimizes CRB through joint transmit beamforming and STAR-RIS coefficient optimization. To further enhance the potential of RIS, stacked intelligent metasurfaces (SIMs), composed of multiple programmable transmissive metasurface layers, extend conventional RIS capabilities by enabling wave-domain processing for ISAC and multiple-input multiple-output (MIMO) precoding~\cite{an2023stacked,zhang2025joint}. However, these designs mainly rely on rigid structures, thereby constraining the spatial degrees of freedom (DoF) available for configuration.

Recently, thanks to developments in micro- and nano-fabrication as well as to the discovery of flexible metamaterials, this progress have enabled the realization of flexible intelligent metasurface (FIM) by depositing dielectric inclusions onto a conformal flexible substrate~\cite{ni2015ultrathin}. These platforms can reshape their supporting surface while manipulating electromagnetic fields, giving rise to adaptive devices that remain effective under changing environmental requirements~\cite{bai2022dynamically}. Recent studies have investigated FIM for multi-target wireless sensing and MIMO communications~\cite{teng2025flexible, an2025flexible}. However, the use of FIM for ISAC remains largely unexplored. Unlike conventional ISAC designs, the morphing of FIM provides an additional degree of design control for ISAC systems. Specifically, FIM implicitly modifies the array steering vectors and the propagation responses seen by both the communication users and the radar receiver. Motivated by this additional DoF, improving the parameter estimation capability in FIM-enabled ISAC systems should be better explored.

To fill that gap, we study the design of FIM-enabled ISAC systems with the objective of minimizing the direction of arrival (DOA) estimation CRB with quality of service (QoS) costraints based on deep reinforcement learning (DRL). More specifically, we formulate an optimization problem to minimize CRB by jointly optimizing the beamforming matrix and the three-dimensional (3D) surface shape of the transmitting and receiving FIMs, subject to QoS requirement as well as the transmit power and deformation constraints. Nevertheless, the problem is highly non-convex, making it difficult to tackle with common gradient-based method. To solve this problem, we utilize a DRL-based framework in which a deterministic actor critic agent is employed to jointly optimize beamforming and FIMs deformation, under a constraint aware reward design that promotes CRB reduction under power, QoS, and surface shape constraints. Simulation results show that our proposed FIM-assisted ISAC design significantly improves CRB optimization performance, compared to conventional rigid array (RA) and other benchmark methods.

\section{System Model}
\begin{figure}
    \centering
    \includegraphics[width=0.80\linewidth]{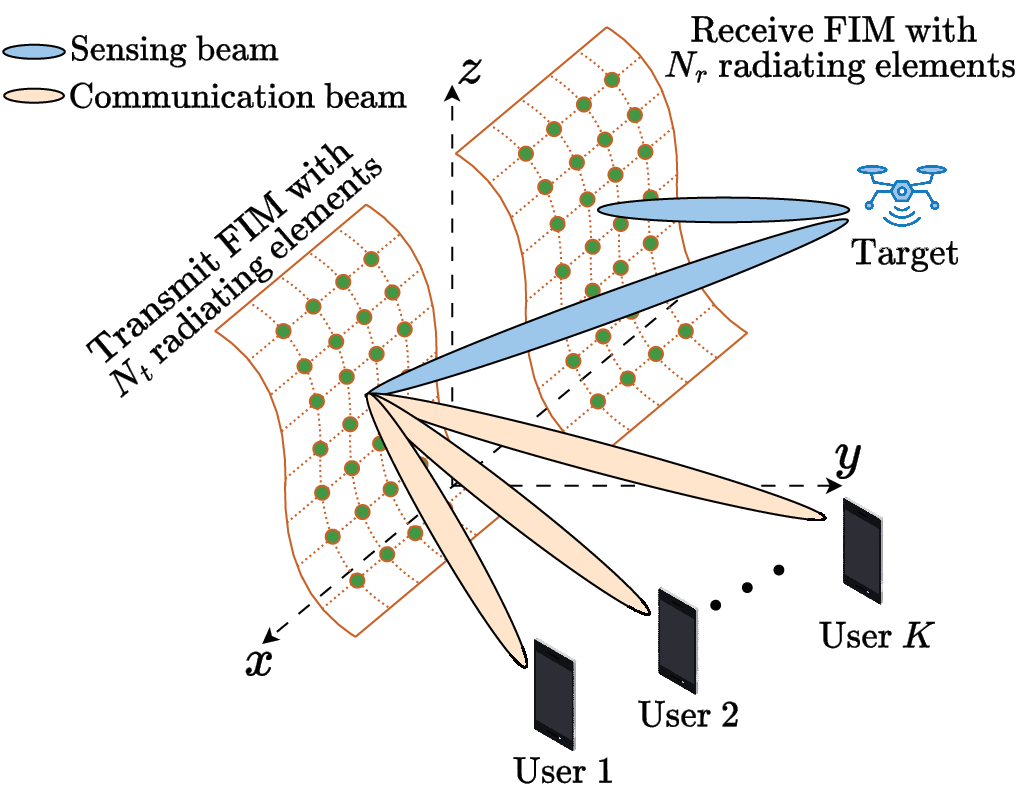}
    \caption{Illustration of the system model.}
    \label{system_model}
    \vspace{-0.3cm}
\end{figure}

We consider an ISAC system aided by a pair of FIMs as shown in Fig.~\ref{system_model}, where the BS is equipped with a transmitting FIM with $N_{\mathrm{t}}$ antennas and a receiving FIM with $N_{\mathrm{r}}$ antennas. FIMs are modeled as a flexible uniform planar array (UPA) on the $x$-$z$ plane. In contrast to a conventional RA, the FIM is able to morph its surface shape~\cite{an2025flexible2}. Specifically, each FIM element can be independently controlled along the $y$-axis, thus we can represent $\bm{p}_{n}^{\mathrm{t}}=[x_{n}^{\mathrm{t}}, y_{n}^{\mathrm{t}}, z_{n}^{\mathrm{t}}]^{\mathrm{T}} \in \mathbb{R}^{3}$, $\forall n \in \mathcal{N}_{\mathrm{t}}$, where $\mathcal{N}_{\mathrm{t}} \triangleq \{ 1,2,\ldots, N_{\mathrm{t}} \}$. Moreover, $\bm{p}_{m}^{\mathrm{r}}=[x_{m}^{\mathrm{r}}, y_{m}^{\mathrm{r}}, z_{m}^{\mathrm{r}}]^{\mathrm{T}} \in \mathbb{R}^{3}$, $\forall m \in \mathcal{N}_{\mathrm{r}}$, where $\mathcal{N}_{\mathrm{r}} \triangleq \{ 1,2,\ldots, N_{\mathrm{r}} \}$, and $\mathbb{R}$ represents the sets of real numbers. Furthermore, the transmit and receive FIMs are arranged as $N^{x}_{\mathrm{t}}\times N^{z}_{\mathrm{t}}$ and $N^{x}_{\mathrm{r}}\times N^{z}_{\mathrm{r}}$ along the $x$-and $z$-axes, respectively. Taking the first receive element as a reference point, we have
\begin{align}
    \label{xn zn}
    x_{m}^{\mathrm{r}} = d_{x} &\times \operatorname{mod}(m-1, N^{x}_{\mathrm{r}}),\\
    z_{m}^{\mathrm{r}} = d_{z} &\times \lfloor (m-1)/N^{x}_{\mathrm{r}} \rfloor,
\end{align}
for the transmit antennas, we represent $x_{n}^{\mathrm{t}} = d_{x} \times \operatorname{mod}(n-1, N^{x}_{\mathrm{t}})+6 \lambda$, and $z_{n}^{\mathrm{t}} = d_{z} \times \lfloor (n-1)/N^{x}_{\mathrm{t}} \rfloor$, where $\lambda$ denotes the wavelength. $d_{x}$ and $d_{z}$ represent the spacing between adjacent antenna elements in the $x$-direction and the $z$-direction, respectively~\cite{teng2025flexible}. We let $\operatorname{mod}(x,y)$ return the remainder from dividing $x$ by $y$, and $\lfloor \cdot \rfloor$ represents the floor function. Furthermore, the $y$-coordinate of each radiating element, i.e., $y^{\mathrm{t}}_{n}$ and $y^{\mathrm{r}}_{m}$, can be adjusted within the maximum range allowed by the reversible deformation of the FIM, it can be written as $y_{\mathrm{min}} \leq y_{n}^{\mathrm{t}} \leq y_{\mathrm{max}}, \forall n \in \mathcal{N}_{\mathrm{t}}$ and $y_{\mathrm{min}} \leq y_{m}^{\mathrm{r}} \leq y_{\mathrm{max}}, \forall m \in \mathcal{N}_{\mathrm{r}}$, respectively. The BS transmits data signals to $K$ communication users with $\mathcal{K} \triangleq \{ 1, \ldots, K \}$, and performs a point-like target detection over $T$ time slots. 

We assume that all channels experience quasi-static flat fading, and $\bm{h}_{k}$ denotes the baseband equivalent channel spanning from the FIM to the receiving antenna of the $k$-th user. For a scatterer in the far field, we denote the elevation angle $\theta \in [0, \pi)$ and the azimuth angle $\phi \in [0, 2 \pi)$, thus the array steering vector $\bm{a}(\bm{y}^{\mathrm{t}}, \phi, \theta)$ of the transmit FIM is given by 
\begin{align}
  \begin{split}
    \label{array steering a}
    \bm{a}(\bm{y}^{\mathrm{t}}, \phi, \theta) &= \left[e^{j\delta(x^{\mathrm{t}}_{1} \cdot \nu + y^{\mathrm{t}}_{1} \cdot \kappa + z^{\mathrm{t}}_{1} \cdot \cos \theta )}, \right. \\
    &\left.  \ldots, e^{j\delta(x^{\mathrm{t}}_{N_{\mathrm{t}}} \cdot \nu + y^{\mathrm{t}}_{N_{\mathrm{t}}} \cdot \kappa + z^{\mathrm{t}}_{N_{\mathrm{t}}} \cdot \cos \theta )} \right]^{\mathrm{T}},
  \end{split}
\end{align}
where $\delta= \frac{2\pi}{\lambda}$, $\bm{y}^{\mathrm{t}}= [y^{\mathrm{t}}_{1}, y^{\mathrm{t}}_{2}, \ldots, y^{\mathrm{t}}_{N_{\mathrm{t}}}]^{\mathrm{T}}$, $\nu = \sin \theta \cdot \cos \phi$, and $\kappa = \sin \theta \cdot \sin \phi$. We let $(\cdot)^{\mathrm T}$ denote the transpose. The elevation angle and azimuth angle of the sensing target are given as $\theta_{\mathrm{t}}$ and $\phi_{\mathrm{t}}$, thus we can get $\bm{a}(\bm{y}^{\mathrm{t}}, \phi_{\mathrm{\mathrm{t}}}, \theta_{\mathrm{t}})$, and the receive FIM array steering is given by
\begin{align}
    \label{receive array steering}
    \bm{b}(\bm{y}^{\mathrm{r}}, \phi_{\mathrm{\mathrm{t}}}, \theta_{\mathrm{t}}) &= \left[e^{j\delta(x^{\mathrm{r}}_{1} \cdot \nu_{\mathrm{t}} + y^{\mathrm{r}}_{1} \cdot \kappa_{\mathrm{t}} + z^{\mathrm{r}}_{1} \cdot \cos \theta_{\mathrm{t}} )}, \right. \\
    &\left.  \ldots, e^{j\delta(x^{\mathrm{r}}_{N_{\mathrm{r}}} \cdot \nu_{\mathrm{t}} + y^{\mathrm{r}}_{N_{\mathrm{r}}} \cdot \kappa_{\mathrm{t}} + z^{\mathrm{r}}_{N_{\mathrm{r}}} \cdot \cos \theta_{\mathrm{t}} )} \right]^{\mathrm{T}},
\end{align}
where $\bm{y}^{\mathrm{r}}= [ y^{\mathrm{r}}_{1}, y^{\mathrm{r}}_{2}, \ldots, y^{\mathrm{r}}_{N_{\mathrm{r}}}]^{\mathrm{T}}$. Furthermore, let $L$ represent the number of propagation paths between the BS and users. Let $\mathbb{C}$ represent the sets of complex numbers. The complex gain of the $\ell$-th path for the $k$-th user is represented by $\alpha_{k,\ell} \in \mathbb{C}, \forall k \in \mathcal{K}$, and $\theta_{\ell}$ and $\phi_{\ell}$ the elevation and azimuth angles at the BS for the $\ell$-th path, respectively. Thus the channel $\bm{h}_{k}$ can be written as
\begin{align}
    \label{com_channel}
    \bm{h}_{k}(\bm{y}^{\mathrm{t}})= \sum_{\ell=1}^{L} \alpha_{k,\ell} \bm{a}(\bm{y}^{\mathrm{t}}, \phi_{\ell}, \theta_{\ell}), \quad \forall k \in \mathcal{K},
\end{align}
where $\alpha_{k,\ell} \sim \mathcal{CN}(0, \rho_{k,\ell}^{2})$, follows the circularly symmetric complex Gaussian  distribution with zero mean and variance $\rho_{k,\ell}^{2}$ corresponding to the average power of the $\ell$-th path for user $k$. Furthermore, $\beta_{k}$ denotes the path loss between the $k$-th user and the BS such that $\sum_{\ell=1}^{L} \rho_{k,\ell}^{2} = \beta_{k}$.

Let $\mathbf{X} \in \mathbb{C}^{N_{\mathrm{t}} \times T}$ be a narrowband ISAC signal matrix as $\mathbf{X} = \mathbf{W}_{\mathrm{C}} \mathbf{S}_{\mathrm{C}} + \mathbf{W}_{\mathrm{R}} \mathbf{S}_{\mathrm{R}} = [\bm{x}[1], \bm{x}[2], \ldots, \bm{x}[T]]$, where $\mathbf{W}_{\mathrm{C}} \in \mathbb{C}^{N_{\mathrm{t}} \times K}$ and $\mathbf{W}_{\mathrm{R}} \in \mathbb{C}^{N_{\mathrm{t}} \times N_{\mathrm{r}}}$ denote the beamforming matrices to be designed for the communication symbols and radar waveforms, respectively. Moreover, $\mathbf{S}_{\mathrm{C}} \in \mathbb{C}^{K \times T}$ contains $K$ unit-power data streams intended for the $K$ users, and $\mathbf{S}_{\mathrm{R}} \in \mathbb{C}^{N_{\mathrm{r}} \times T}$ is the dedicated sensing signal. Furthermore, we can define the beamforming matrix $\mathbf{W} \triangleq [\mathbf{W}_\mathrm{C}~ \mathbf{W}_\mathrm{R}] \in \mathbb{C}^{N_{\mathrm{t}} \times (K+N_{\mathrm{r}})}$ and $\mathbf{S} \triangleq [\mathbf{S}_{\mathrm{C}}^\mathrm{T} ~ \mathbf{S}^{\mathrm{T}}_{\mathrm{R}}]^{\mathrm{T}}$. The data streams are assumed to be independent with each other so that $\lim\limits_{T \to \infty}\frac{1}{T}\mathbf{S}\mathbf{S}^{\mathrm{H}} = \mathbf{I}_{K+N_{\mathrm{r}}}$, where $\mathbf{I}_{L}$ denotes an identical matrix of size $L \times L$. The received signal at $k$-th user in the $t$-th time slot is given by
\begin{align}
    \label{com_channel}
    y_{k}[t] = \bm{h}_{k}^{\mathrm{H}}(\bm{y}^{\mathrm{t}})\bm{x}[t] + n_{k},
\end{align}
where $n_{k} \sim \mathcal{CN}(0,\sigma_{k}^{2})$ is the additive white Gaussian noise (AWGN) at the $k$-th user, with $\sigma_{k}^{2}$ representing the average noise power.

By transmitting $\mathbf{X}$ to sense a target, the reflected echo signal matrix at the receiver of the BS is given by $\mathbf{Y}_{\mathrm{R}} = \mathbf{G} \mathbf{X} + \mathbf{N}_{\mathrm{R}}$, where $\mathbf{N}_{\mathrm{R}} \in \mathbb{C}^{N_{\mathrm{r}} \times T}$ denotes the AWGN matrix, with variance of each entry being $\sigma_{\mathrm{r}}^{2}$, and $\mathbf{G} \in \mathbb{C}^{N_{\mathrm{r}} \times N_{\mathrm{t}}}$ represents the target response matrix. In our work, the target is modeled as an unstructured point that is far away
from the BS. The target response matrix can be written as
\begin{align}
    \label{target response matrix}
    \mathbf{G}= \alpha_{\mathrm{r}} \bm{b}(\bm{y}^{\mathrm{r}}, \phi_{\mathrm{t}}, \theta_{\mathrm{t}}) \bm{a}^{\mathrm{H}}(\bm{y}^{\mathrm{t}},\phi_{\mathrm{t}}, \theta_{\mathrm{t}}),
\end{align}
where $\alpha_{\mathrm{r}}$ denotes the complex-valued channel coefficient that depends on the target radar cross section (RCS) and the round-trip path loss, and $(\cdot)^{\mathrm H}$ denotes the conjugate transpose. To drive the CRB for estimating angles, we first vectorize the received signal $\mathbf{Y}_{\mathrm{R}}$ as
\begin{align}
    \label{vec yr}
    \tilde{\bm{y}}^{}_{\mathrm{R}} = \text{vec} (\alpha_{\mathrm{r}} \bm{b}(\bm{y}^{\mathrm{r}}, \phi_{\mathrm{t}}, \theta_{\mathrm{t}}) \bm{a}^{\mathrm{H}}(\bm{y}^{\mathrm{t}},\phi_{\mathrm{t}}, \theta_{\mathrm{t}})\mathbf{W}\mathbf{S}) + \bm{n}_{\mathrm{r}},
\end{align}
where $\bm{n}_{\mathrm{r}} \triangleq \text{vec}(\mathbf{N}_{\mathrm{R}})$, and vec$({\cdot})$ denotes the column-wise vectorization of the a matrix. We let $\bm{\xi} = [\bm{\theta}, \bm{\alpha}]^{\mathrm{T}}$ denote the vector of four unknown real parameters to be estimated, $\bm{\theta} = [\phi_{\mathrm{t}}, \theta_{\mathrm{t}}]$, and $\bm{\alpha} = [\text{Re}\{\alpha_{\mathrm{r}}\}, \text{Im}\{ \alpha_{\mathrm{r}}\}]$. Furthermore, we can obtain the CRB matrix from the inverse of the Fisher information matrix. Specifically, we abbreviate $\bm{b}(\bm{y}^{\mathrm{r}}, \phi_{\mathrm{t}}, \theta_{\mathrm{t}})$ as $\bm{b}$ and $\bm{a}(\bm{y}^{\mathrm{t}},\phi_{\mathrm{t}}, \theta_{\mathrm{t}})$ as $\bm{a}$, and the $(q,\zeta)$-th element of the Fisher information matrix $\mathbf{F} \in \mathbb{C}^{4 \times 4}$ is given by
\begin{equation}
  \begin{aligned}
    \label{Fisher inf matrix}
    \mathbf{F}(q,\zeta) &= \frac{2}{\sigma^{2}_{\mathrm{r}}} \mathrm{Re} \left\{ \frac{\partial{\alpha_{\mathrm{r}}}\text{vec}(\bm{b}\bm{a}^{\mathrm{H}}\mathbf{W}\mathbf{S})^{\mathrm{H}}}{\partial{\xi_{q}}} \frac{\partial{\alpha_{\mathrm{r}}}\text{vec}(\bm{b}\bm{a}^{\mathrm{H}}\mathbf{W}\mathbf{S})}{\partial{\xi_{\zeta}}} \right\}\\
    &= \frac{2}{\sigma^{2}_{\mathrm{r}}} \mathrm{Re} \left\{ \frac{\partial{\alpha_{\mathrm{r}}}\text{vec}(\mathbf{A}\mathbf{W}\mathbf{S})^{\mathrm{H}}}{\partial{\xi_{q}}} \frac{\partial{\alpha_{\mathrm{r}}}\text{vec}(\mathbf{A}\mathbf{W}\mathbf{S})}{\partial{\xi_{\zeta}}} \right\},
  \end{aligned}
\end{equation}
where
\begin{align}
    \label{partial 1}
    \frac{\partial\alpha_{\mathrm{r}}\text{vec}\left( \mathbf{AWS} \right)}{\partial \theta_{1}}&=\alpha_{\mathrm{r}} \text{vec}\left( \mathbf{\dot{A}WS} \right),\\
    \label{partial 2}
    \frac{\partial\alpha_{\mathrm{r}}\text{vec}\left( \mathbf{AWS} \right)}{\partial \theta_{2}}&=\alpha_{\mathrm{r}} \text{vec}\left( \mathbf{\ddot{A}WS} \right),\\
    \label{partial 3}
    \frac{\partial\alpha_{\mathrm{r}}\text{vec}\left( \mathbf{AWS} \right)}{\partial \bm{\alpha}}&=[1 \ j]^{\mathrm{T}} \otimes \text{vec}\left( \mathbf{AWS} \right),
\end{align}
where ${\mathbf{\dot{A}}}$ and ${\mathbf{\ddot{A}}}$ denote the partial derivatives of $\mathbf{A}$ with respective to $\phi_{\mathrm{t}}$ and $\theta_{\mathrm{t}}$, respectively. We let $\partial(\cdot)$ denote the partial differential of a function, and $\otimes$ denotes the Kronecker product. Thus, plugging~\eqref{partial 1} --\eqref{partial 3} into~\eqref{Fisher inf matrix}, the element of Fisher information matrix can be calculated as
\begin{align}
    \label{phiphi}
    F_{\phi_{\mathrm{t}}, \phi_{\mathrm{t}}} &= \frac{2T\left| \alpha_{\mathrm{r}} \right|^{2}}{\sigma^{2}_{\mathrm{r}}} \text{Re}\left\{ \text{Tr}\left\{ \mathbf{\dot{A}WW^{\mathrm{H}}\dot{A}^{\mathrm{H}}}  \right\} \right\},\\
    \label{Ftheta alpha}
     \mathbf{F}_{\theta_{1}, \bm\alpha^{\mathrm{T}}} &= \frac{2T}{\sigma^{2}_{\mathrm{r}}} \text{Re}\left\{ \text{Tr}\left\{ \alpha_{\mathrm{r}}^{*}\mathbf{AWW^{\mathrm{H}}\dot{A}^{\mathrm{H}}}  \right\} [1 \ j]\right\},\\
    \label{alpha alpha}
    \mathbf{F}_{\bm{\alpha}, \bm\alpha^{\mathrm{T}}} &= \frac{2T}{\sigma^{2}_{\mathrm{r}}} \text{Tr}\left\{ \mathbf{AWW^{\mathrm{H}}A^{\mathrm{H}}}  \right\} \mathbf{I}_{2}.
\end{align}

Then, we can also get $ F_{\phi_{\mathrm{t}}, \theta_{\mathrm{t}}}$ and $F_{\theta_{\mathrm{t}}, \theta_{\mathrm{t}}}$, $\mathbf{F}_{\theta_{2}, \bm\alpha^{\mathrm{T}}}$, and we have $F_{\theta_{\mathrm{t}}, \phi_{\mathrm{t}}} = F_{\theta_{1},\theta_{2}}$. Thus the sub-matrices of $\mathbf{F}$ can be written as
\begin{align}
    \label{Fthetatheta^T}
    \mathbf{F}_{\bm{\theta}, \bm{\theta}^{\mathrm{T}}} &= 
    \begin{bmatrix}
        F_{\phi_{\mathrm{t}}, \phi_{\mathrm{t}}} &
        F_{\phi_{\mathrm{t}}, \theta_{\mathrm{t}}}\\
        F_{\theta_{\mathrm{t}}, \phi_{\mathrm{t}}} &
        F_{\theta_{\mathrm{t}}, \theta_{\mathrm{t}}}
    \end{bmatrix},\\
    \label{Fthetaalpha}
    \mathbf{F}_{\bm{\theta}, \bm{\alpha}^{\mathrm{T}}} &= 
    \begin{bmatrix}
        F_{\phi_{\mathrm{t}}, \bm{\alpha}^{\mathrm{T}}}\\
        F_{\theta_{\mathrm{t}}, \bm{\alpha}^{\mathrm{T}}}
    \end{bmatrix}.
\end{align}

The CRB matrix $\mathbf{C}$ is the inverse of $\mathbf{F}$ and the diagonal elements of $\mathbf{C}$ represent the CRB for $\bm\xi$. To derive CRB for DOA estimation, we partition $\mathbf{F}$ and $\mathbf{C}$ into $2 \times2$ blocks as 
\begin{align}
    \label{C matrix = F}
    \mathbf{C} = 
    \begin{bmatrix}
        \mathbf{C}_{\bm{\theta}\bm{\theta}^{\mathrm{T}}} & \mathbf{C}_{\bm{\theta}\bm{\alpha}^{\mathrm{T}}} \\
        \mathbf{C}_{\bm{\alpha}\bm{\theta}^{\mathrm{T}}} &
        \mathbf{C}_{\bm{\alpha}\bm{\alpha}^{\mathrm{T}}}
    \end{bmatrix}
    =
    \begin{bmatrix}
        \mathbf{F}_{\bm{\theta}\bm{\theta}^{\mathrm{T}}} & \mathbf{F}_{\bm{\theta}\bm{\alpha}^{\mathrm{T}}} \\
        \mathbf{F}^{\mathrm{T}}_{\bm{\theta}\bm{\alpha}^{\mathrm{T}}} &
        \mathbf{F}_{\bm{\alpha}\bm{\alpha}^{\mathrm{T}}}
    \end{bmatrix}^{-1}
    = \mathbf{F}^{-1}.
\end{align}

The CRB for estimating $\bm{\theta}$ can be obtained as  \vspace{-0.2cm}
\begin{equation*}
    \label{crb1+crb2}
    \text{CRB}_{\theta_{1}} + \text{CRB}_{\theta_{2}} = \text{Tr} \left\{ \left( \mathbf{F}_{\bm{\theta}\bm{\theta}^{\mathrm{T}}} -  \mathbf{F}_{\bm{\theta}\bm{\alpha}^{\mathrm{T}}}   \mathbf{F}^{-1}_{\bm{\alpha}\bm{\alpha}^{\mathrm{T}}}  \mathbf{F}^{\mathrm{T}}_{\bm{\theta}\bm{\alpha}^{\mathrm{T}}}  \right)^{-1} \right\}.
\end{equation*}

\vspace{-0.35cm}
\section{CRB Minimization Problem Based on DRL}
\subsection{Problem Formulation}
We aim to minimize the CRB by jointly optimizing the transmit beamforming matrix, and the surface shapes of two FIMs within the morphing range $y_{\mathrm{max}}$ and $y_{\mathrm{min}}$ under the total transmit power $P_{\mathrm{max}}$ and the QoS constraints. The CRB minimization problem is given by \vspace{-0.15cm}
\begin{subequations}\label{form opt problem}
  \begin{align}
      \mathop {\min}\limits_{\mathbf{W}, \bm{y}^{\mathrm{t}}, \bm{y}^{\mathrm{r}}} \quad& \text{Tr} \left\{ \left( \mathbf{F}_{\bm{\theta}\bm{\theta}^{\mathrm{T}}} -  \mathbf{F}_{\bm{\theta}\bm{\alpha}^{\mathrm{T}}}   \mathbf{F}^{-1}_{\bm{\alpha}\bm{\alpha}^{\mathrm{T}}}  \mathbf{F}^{\mathrm{T}}_{\bm{\theta}\bm{\alpha}^{\mathrm{T}}}  \right)^{-1} \right\} \label{main problem} \\
      \text{s.t.} \qquad& \left\|\mathbf{W}\right\|^{2}_{\mathrm{F}} \leq P_{\mathrm{max}}, \label{s.t. W} \\
      & \frac{\left|\bm{h}_{k}^{\mathrm{H}} \bm{w}_{k}\right|^{2}}{\sum_{i=1,i\neq k}^{K+N_{\mathrm{r}}}\left|\bm{h}_{k}^{\mathrm{H}} \bm{w}_{i}\right|^{2}+\sigma_{k}^{2}} \geq \gamma, \label{s.t. r_th} \\
      & y_{\mathrm{min}} \leq  y^{\mathrm{t}}_{n} \leq y_{\mathrm{max}}, \label{s.t. yt}\\
      & y_{\mathrm{min}} \leq  y^{\mathrm{r}}_{m} \leq y_{\mathrm{max}},\label{s.t. yr}
  \end{align}
\end{subequations}
where \eqref{s.t. W} means that the total transmit power is bounded by a maximum value $P_{\mathrm{max}}$, \eqref{s.t. r_th} indicates that signal-to-interference-plus-noise ratio (SINR) of the $k$-th user is no less than the threshold $\gamma$, and $\mathbf{W} = [\bm{w}_{1}, \bm{w}_{2}, \ldots, \bm{w}_{K+N_{\mathrm{r}}}]$. \eqref{s.t. yt} and \eqref{s.t. yr} represent the maximum deformation range of FIMs.

\vspace{-0.3cm}
\subsection{DDPG Algorithm for the Optimization Problem}
The joint optimization of the beamforming matrix and FIMs surface shape is a highly non-convex problem and is difficult to solve. To address this challenge, we utilize the deep deterministic policy gradient (DDPG) algorithm to solve it. The algorithm contains $D$ episodes, each episode has $E$ steps, and at each step, the agent can choose an action $\Lambda^{(e)} \in \mathcal{A}$ according to the current state $\psi^{(e)} \in \mathcal{S}$, resulting in the reward $\Gamma^{(e)}$ and the next state $\psi^{(e+1)}$, where $\mathcal{A}$ and $\mathcal{S}$ denote the action space and state space, respectively. Specifically, the state space, action space, and reward function are defined as follows.
\begin{enumerate}
	\item State: At step $e$, state $\psi_{e}$ is determined by the BS transmit power, the antenna positions of transmit and receive FIMs, the $k$-th user SINR and CRB values, the power ratio of the user $k$, and matrix $\mathbf{W}$. Furthermore, we define three violation indicators that are used to quantify breaches of power, QoS, and position constraints. Thus, the dimension of state space is $7+2K+N_{\mathrm{r}}+N_{\mathrm{t}}$.
	\item Action: The action consists of the first $2K$ dimensions, which determine the power allocation among $K$ users to satisfy the power and SINR constraints, as well as control the beam direction of the $k$-th user. In addition, we use a global allocation factor and a perceived direction factor to determine the global power partition between communication and sensing, meanwhile balance the sensing interference. Finally, by uniformly thresholding continuous $[-1,1]$ outputs, we obtain discrete antenna position choices and adjust the $N_{\mathrm{t}}$ transmit and $N_{\mathrm{r}}$ receive antenna positions. Therefore, the dimension of the action space is $2+2K+N_{\mathrm{t}}+N_{\mathrm{r}}$.
	\item Reward: The reward is determined by the value of $\mathrm{CRB}_{\theta_{1}}+\mathrm{CRB}_{\theta_{2}}$, in order to ensure the QoS constraint, we added a penalty for each user who does not meet the SINR constraint, the reward function can be given as
	\begin{align}
		\label{reward function}
		\Gamma^{(e)} = f(\mathrm{CRB}^{e}_{\theta_{1}}+\mathrm{CRB}^{e}_{\theta_{2}}) - \varpi \sum_{k=1}^{K}\left| \varsigma_{k,e} - \gamma \right|, 
	\end{align}
	where $f(\cdot)$ denotes a log-domain normalized CRB function that maps the current CRB value at step $e$ to a bounded reward using the target and worst reference values, $\varpi$ represents the penalty weight for QoS constraint, and $\varsigma_{k,e}$ denotes the SINR of user $k$ at step $e$.
\end{enumerate}

The DDPG network is based on actor critic architecture, at each training step, we uniformly sample a mini-batch of size from the replay buffer and use it to update both the critic and actor networks. The update for the critic network is given as
\begin{align}
    \label{critic network parameter upate}
    \vartheta_{c}^{(e+1)} &= \vartheta_{c}^{(e)}-\mu_{c}\nabla_{\vartheta_{c}} \varrho\left( \vartheta_{c} \right),\\
    \label{loss function}
    \begin{split}
    \varrho\left( \vartheta_{c} \right) &=
    \left( \Gamma^{(e)} + \varUpsilon Q\left( \vartheta_{c}^{(\text{target})} \mid  \psi^{(e+1)} , \Lambda^{\prime}  \right) \right.\\
    &\left. \quad - Q\left( \vartheta_{c} \mid \psi^{(e)} , \Lambda^{(e)} \right)  \right)^{2},
    \end{split}
\end{align}
where $\mu_{c}$ denotes the learning rate for training the Critic network update, $\Lambda^{\prime}$ denotes the action output in the next state $\psi^{(e+1)}$, $\vartheta_{c}^{(\text{target})}$ is the target network parameter, and $\varUpsilon$ represents the discount factor. \eqref{loss function} defines the loss function of the critic network, which is constructed as minimizing the squared error between the current Q value $Q\left( \vartheta_{c} \mid \psi^{(e)} , \Lambda^{(e)} \right)$ and the target Q value $Q\left( \vartheta_{c}^{(\text{target})} \mid  \psi^{(e+1)} , \Lambda^{\prime}  \right)$, while~\eqref{critic network parameter upate} updates the critic parameters by minimizing this loss through gradient descent. $Q(\cdot)$ is constructed based on the Bellman equation. The update process of the actor network is given by
\begin{align}
  \begin{split}
    \label{actor update}
    \vartheta_{o}^{(e+1)} &= \vartheta_{o}^{(e)}- \mu_{o}\nabla_{o} Q\left( \vartheta_{c}^{(\text{target})} \mid  \psi^{(e)} , \Lambda  \right) \\
    & \quad \cdot \nabla_{\vartheta_{o}} \pi\left( \vartheta_{o} \mid \psi^{(e)} \right),
  \end{split}
\end{align}
where $\mu_{o}$ denotes the learning rate of the actor network, and $\pi\left( \vartheta_{o} \mid \psi^{(e)} \right)$ represents the deterministic policy parameterized by $\vartheta$, which maps the current state $\psi^{(e)}$ to the action $\Gamma^{(e)}$. \eqref{actor update} means that by applying the deterministic policy gradient, updating the actor parameters to maximize the critic estimated Q value, thereby promoting actions with a higher expected return. Finally, the target networks is given as
\begin{align}
    \label{target critic network and the target actor network update}
    \vartheta_{c}^{\text{target}} &\leftarrow \tau_{c}\vartheta_{c}+(1-\tau_{c})\vartheta_{c}^{\text{target}},\\
    \label{second target critic network and the target actor network update}
    \vartheta_{o}^{\text{target}} &\leftarrow \tau_{o}\vartheta_{o}+(1-\tau_{o})\vartheta_{o}^{\text{target}},
\end{align}
where $\tau_{c}$ and $\tau_{o}$ are the learning rate of the target critic network and the target actor network, respectively.

\section{Simluation Results}
\begin{figure}[t]
    \centering
    \includegraphics[width=0.75\linewidth]{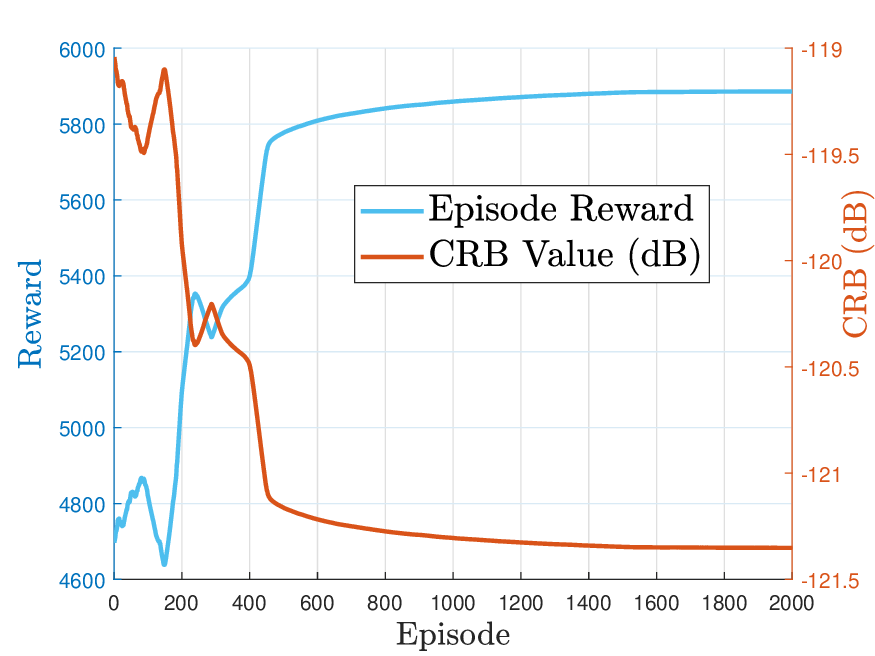}
    \vspace{-0.2cm}
    \caption{CRB and reward for $\mu_{c} = 3\times10^{-4}$, $\mu_{o} = 3\times10^{-5}$, $K=4$.}
    \label{lr_crb}
    \vspace{-0.5cm}
\end{figure}
\begin{figure}[t]
    \centering
    \includegraphics[width=0.75\linewidth]{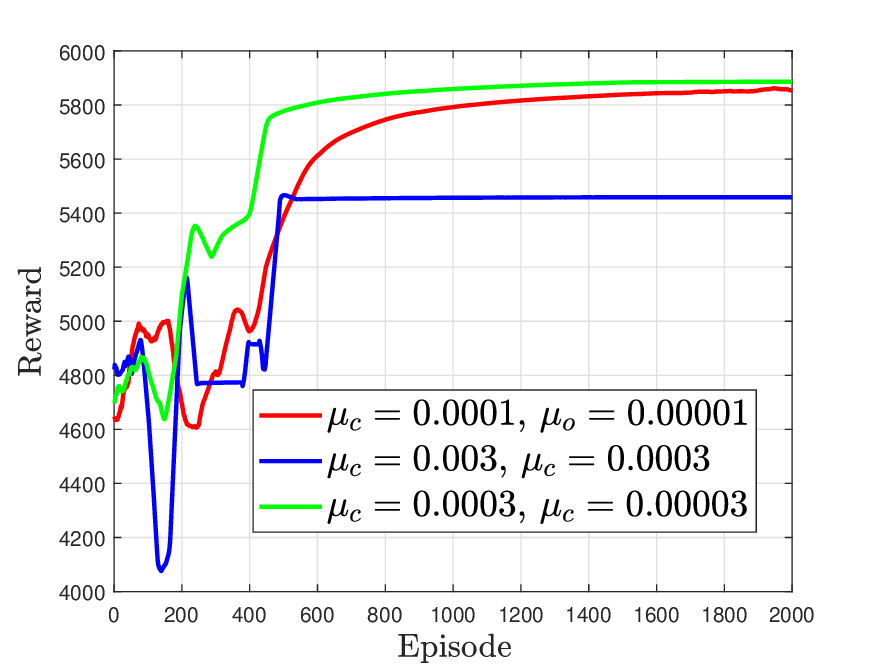}
    \vspace{-0.2cm}
    \caption{Reward under different learning rates for $K=4$, $\gamma=2^{5}-1$.}
    \label{diff_lr}
    \vspace{-0.5cm}
\end{figure}
\begin{figure}[t]
    \centering
    \includegraphics[width=0.75\linewidth]{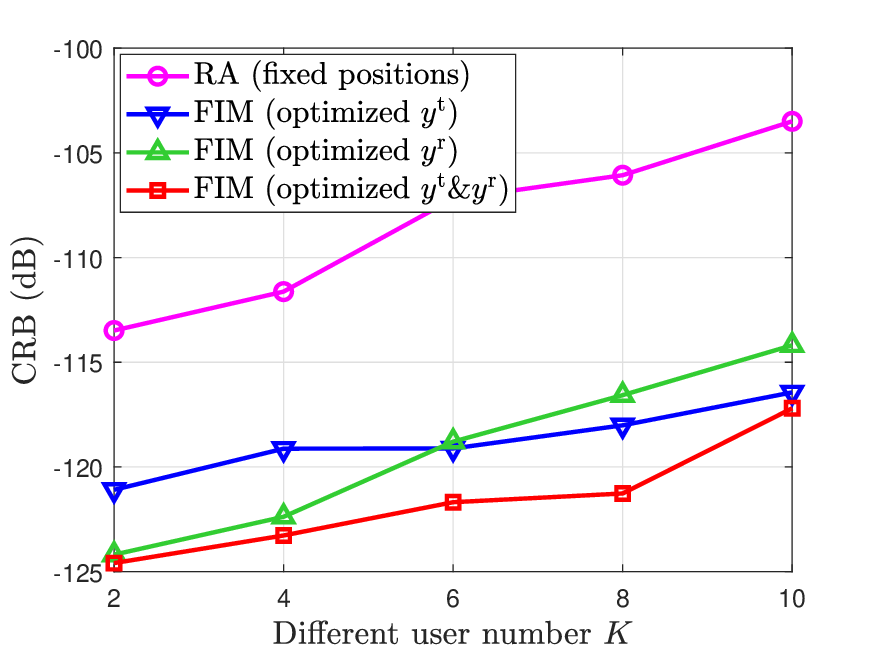}
    \vspace{-0.2cm}
    \caption{CRB under different users for $P_{\mathrm{max}}=20$ dBm, $\gamma=2^{3}-1$, $\mu_{c} = 3\times10^{-4}$, $\mu_{o} = 3\times10^{-5}$.}
    \label{d_users}
    \vspace{-0.5cm}
\end{figure}
\begin{figure}[t]
    \centering
    \includegraphics[width=0.75\linewidth]{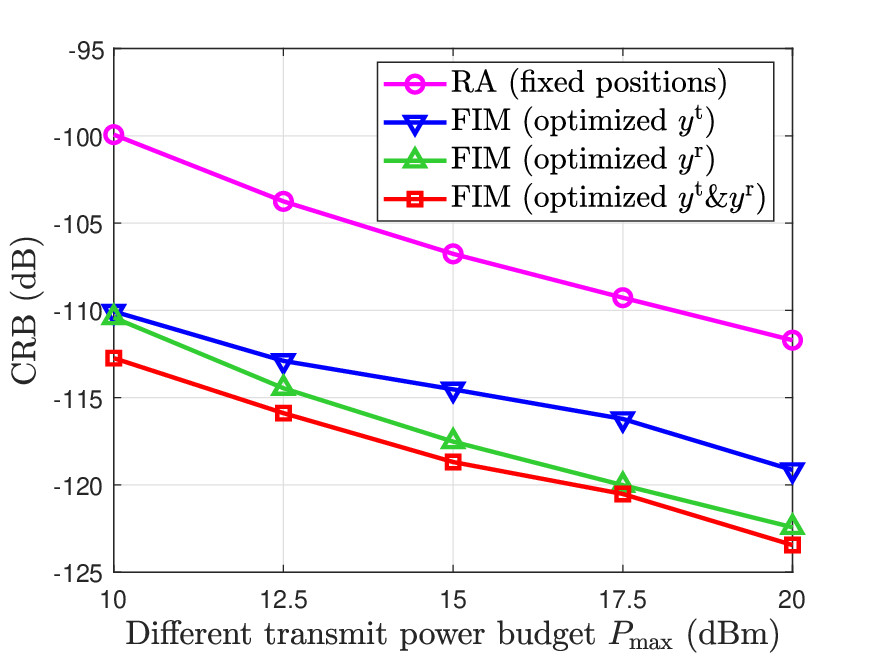}
    \caption{CRB under different $P_{\mathrm{max}}$ with $K=4$, $\gamma=2^{3}-1$, $\mu_{c} = 3\times10^{-4}$, $\mu_{o} = 3\times10^{-5}$.}
    \label{d_Pmax}
    \vspace{-0.5cm}
\end{figure}
\begin{figure}[t]
    \centering
    \includegraphics[width=0.75\linewidth]{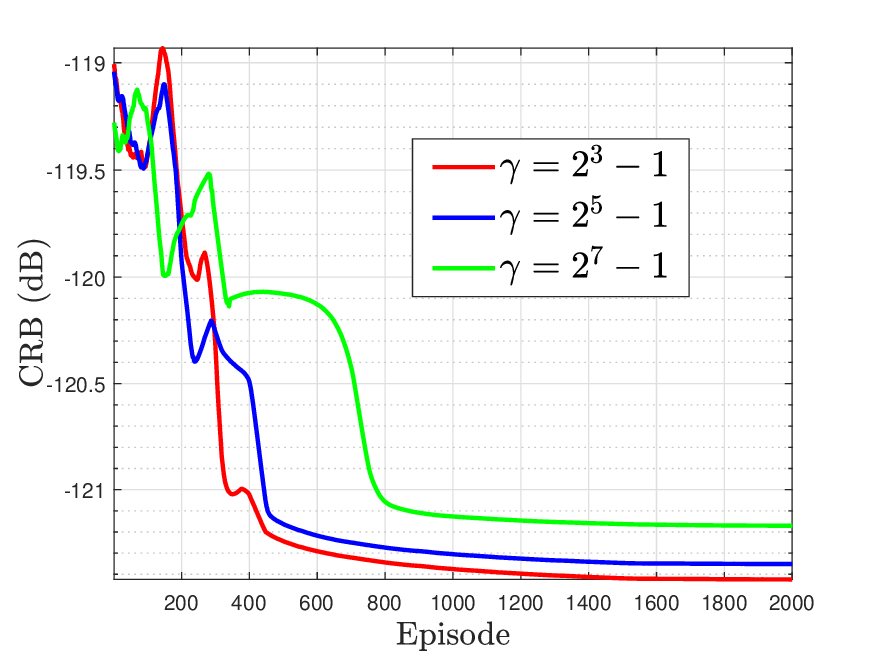}
    \vspace{-0.2cm}
    \caption{CRB convergence under different thresholds for $K=4$, $P_{\mathrm{max}}=20$ dBm, $\mu_{c} = 3\times10^{-4}$, $\mu_{o} = 3\times10^{-5}$.}
    \label{d_gamma}
    \vspace{-0.4cm}
\end{figure}

In this section, simulations are carried out to evaluate the proposed DRL method for jointly optimizing the FIMs and the beamforming matrix. The number of transmit and receive elements of FIMs are $N_{\mathrm{t}} = 9$ and $N_{\mathrm{r}} = 12$, respectively. The spacing between adjacent elements along the $x$- and $z$-axes of FIMs is $0.5 \lambda$, and the deformation range of the FIMs is $y_{\mathrm{max}}=-y_{\mathrm{min}}=\lambda$. The range of each deformation step is $\lambda/8$. There is a sensing target and $K$ users. $\sigma_{k}^{2} = -100$ dBm, and $\sigma_{\mathrm{r}}^{2} = -95$ dBm. The actor and critic networks are two hidden layer networks with 256 units per layer. The discount factor $\varUpsilon$ is $0.99$, $\tau_{c}=\tau_{o}=0.002$, the buffer size is $100{,}000$. We set the number of episode $D$ and mini-batch size as $2{,}000$ and $128$, respectively. There are $800$ steps of each episode. The plotted CRB values are moving averages over 50 episodes.

As shown in Fig.~\ref{lr_crb}, the CRB decreases and stabilizes over training, while the reward increases and converges. This trend indicates a stable training process and demonstrates that the designed reward is well aligned with the CRB minimization objective, enabling the agent progressively reduces the CRB.

We compare different learning rate when minimizing the CRB in Fig.~\ref{diff_lr}, and use a smaller learning rate for the actor as in~\cite{henderson2018deep}. We can see that the learning rates for $\mu_{c} =3\times10^{-3}$ and $\mu_{o} =3\times10^{-4}$ result in the worse performance, because too large learning rate can increase training oscillations, which cause a sharp degradation in performance. Meanwhile, lower learning rates lead to slow convergence and poor performance. 

We evaluate the impact of different users for CRB in Fig.~\ref{d_users}. Compared with RA baseline, the transmit FIM shape optimization baseline, and the receive FIM shape optimization baseline, our proposed FIM-based surface shape optimization scheme, which jointly optimizes the transmit and receive FIMs shape, can significantly reduce the CRB. Furthermore, with an increasing number of users, stronger multiuser interference and stricter QoS constraints shrink the feasible beamforming, reducing the Fisher information for estimation.

It can be observed from Fig.~\ref{d_Pmax} that the CRB achieved by all schemes decreases as $P_{\mathrm{max}}$ increases, since higher transmit power improves the echo signal-to-noise ratio, thus increasing Fisher information. Furthermore, our FIM-assisted method consistently achieves a lower CRB than the other methods.

We compare the CRB curves over episodes under different threshold values in Fig.~\ref{d_gamma}, the results indicate that our DRL-based method can achieve convergence under different thresholds, and the CRB increases with higher threshold $\gamma$. Because more beamforming and power resources must be allocated to meet QoS, then reducing the sensing Fisher information.

\vspace{-0.2cm}
\section{Conclusion}
In this work, we investigated an FIM-aided ISAC system, where the transmit and receive FIMs could adaptively morph their surface shape to minimize the CRB. To achieve this, we formulated a problem that jointly optimizes the FIMs surface shape and the beamforming matrix. Due to the non-convex of the optimization problem, we utilize a deterministic actor critic DRL-based framework that incorporates constraints to drive CRB reduction. The results demonstrated that our joint transmit–receive FIMs surface shaping
approach outperforms the RA and other baselines in terms of CRB. In future work, we will investigate more advanced wideband scenarios to enhance system applicability.

\vspace{-0.2cm}
\bibliography{ckwx}

\end{document}